# A criterion for bubble merging in liquid metal: computational and experimental study

Mojtaba Barzegari[a,*], Hossein Bayani[b], Seyyed Mohammad Hossein Mirbagheri[b]

[a] Department of Biomedical Engineering, Faculty of New Sciences and Technologies, University of Tehran, Tehran, Iran

[b] Department of Mining and Metallurgical Engineering, Amirkabir University of Technology, Tehran, Iran

[*] Corresponding author: P.O.B. 1439957131 Tehran, Iran, mbarzegary@alumni.ut.ac.ir

## 1. Abstract

An innovative model is presented for merging of bubbles inside a liquid metal. The proposed model is based on forming a thin film (narrow channel) between merging bubbles during growth. Rupturing of the film occurs when an oscillation in velocity and pressure arises inside the channel followed by merging of the bubbles. The proposed model –based on lattice Boltzmann Method - is capable of simulating merging bubbles in micro, meso, and macro-scales with no limitation on the number of bubbles. Experimental studies reveal a good consistency between modeling results and real conditions.

## 2. Introduction

Metal foams as a complex of solid bubbles are known for their unique physical and mechanical properties. It is generally accepted that formability of metal foams is intimately correlated with the presence of particles. The physical understanding of foam formation in the presence of colloidal particles with no surface active material is very complex and still rather poor [1]. A variety of studies has been done to analyze the bubble stabilization parameters, most of which are focused on ionic liquids, especially in water. In addition, due to the presence of metallic bond in metal melt, no ionic or polar attraction/repulsion force is present, causing a different behavior of the liquid-gas interface in molten-metal in comparison with aqueous solutions. Different approaches are available for bubbling simulation with their own advantages and disadvantages. Lattice Boltzmann Method (LBM) is the most accurate numerical method. It models the microscopic and mesoscopic kinetic equations and requires no equation or correlation to include source terms in governing equations. Also interface is a post-processed quantity and not a mathematical boundary [2]. Bermond et al performed computational investigation on bubble interactions [3]. Some researchers used LBM to simulate fluid flow to determine bubble coalescence conditions [4] .

In this study we propose a new model based on Shan-Chen scheme [5] for prediction of bubbles´ merging phenomena in Al-Si liquid foams without kiss-point rupturing. Simulation results were validated through conducting experimental work.

## 3. Mathematics and Physical models:

When foaming starts, bubbles are very small (nuclei bubbles < 5 microns) and are abruptly produced by dissolution reaction of blowing agent around the melt (micro-scale). Then growth (meso-scale, above 100 microns) starts and bubbles' equilibrium throughout the melt is a function of various underlying forces. Bubble growth, coalescence by cell wall rupture, drainage, and finally foam collapse are the four main stages of the process (macro-scale). Fluid dynamics, interfacial and capillary forces are factors behind the scene [1]. There are two phases in a foam melt system: gas and liquid. Interaction of these two phases forms the final structure of the foam. Interactions on phase boundaries and choice of boundary conditions is deterministic. By using a random function, bubble growth starts after nucleation. To model the growth phenomena in each time step, gas is added by virtual blowing agent to each bubble, giving rise to bubble volume due to pressure balance. Gas blowing (growth) continues until blowing agent is finished. Other phenomena like drainage and wall rupture are considered in growth process by equilibrium bubble equations. Wall rupture and merging phenomenon need additional criterion which could be either an experimental one or a mathematical condition .

Two mechanisms are proposed for bubble merging. Oswald Ripening based on the diffusion of gas from small bubbles to the bigger ones and film rupture [6] for which, two mechanisms are proposed; i.e. void nucleation and growth (surface tension) and thin film instability (surface perturbations). The former deals with random formation of micro-bubbles in the thin film and consumes energy ($E_r$). A critical radius ($r^*$) is needed for void stability below which voids would disappear. This, requires activation energy $E_a=E(r^*)$ to start described by Arrhenius equation [6]. In the second mechanism, if the film thickness falls below a critical value due to drainage and merging, a perturbation occurs. The instability in the thin film appears when the vibration wavelength exceeds a critical value ($\lambda_c$), leading to film rupture. The thin film can be considered as a disc with radius $R_d$ and thickness h enclosed by two interfaces with surface tension $\gamma$ (Fig. 1). Then the critical thickness of merging is $h_c=k\,(R_d/\gamma^{0.5})^{0.5}$ with $k$ being a constant. Ignoring this limitation, thickening would continue up to the formation of a molecular wall. But in the presence of stabilizing particles, surfactants or oxide films, the situation is different. Inclusions trigger another mechanism. A novel approach for determination of the onset of wall rupture using the second derivative of the pressure ($\nabla^2 p=0$) is employed based on which, the critical wall thickness ($h_c$) is calculated.

### 3.1. Numerical Model

A time random model is developed to determine nucleation site positions. Decomposition of the blowing agents is accordingly calculated for these positions. Other lattice points are filled with a liquid with A356 alloy density. All parameters are changed to dimensionless properties by OpenLB code [7]. Shan-Chen model is based on incorporating long-range attractive forces (F) between distribution functions. Based on classical mechanics theory, the potential force is defined as $p_0 = \rho c_s^2 + 0.166 G \psi^2$ in which $\rho$, $c_s$, and G are density, speed of sound, and strength of the attraction respectively. This equation can describe the separation of phases by choosing suitable pseudo-potential function $\psi(x)$. In the case of Al-Si alloy liquid and the hydrogen gas, the density ratio is large. Hence, we chose the pseudo-potential as $\psi(\rho) = 1 - \exp(-\rho)$. This allows separation of gas and liquid in larger density ratios [8].

Present model is a modification of Shan-Chen model in which, each phase is considered in a separate lattice (grid), i.e. two separate simulations are carried out simultaneously for a two-phase model. In Fig. 1, the situation is different when an overlap occurs. In the first stage of time step, when applying streams and collisions, the corona of the bubbles contact each other, while there is no interference in the fluid domain. The interference causes bubbles directly impact each other. This is in contradiction with the fact that each interaction is only performed through the fluid domain. This leads to the spontaneous formation of the barrier wall and the involved bubbles will never merge. This innovative computational process is presented schematically in Fig. 1.

Now we need a criterion for determination of the barrier wall decay time (thin film rupture). In presented model, we use the second derivative of the pressure profile across the interaction zone to disable the above algorithm and let the bubbles merge (i.e. $\eta_m=1$; if $\partial^2 p/\partial n^2 \neq 0$ and $\eta_m=0$; if $\partial^2 p/\partial n^2=0$). By applying this condition, bubbles merge rapidly and their dynamic effect on liquid domain becomes observable.

### 3.2. Experimental model:

Aluminum A356 alloy was melted with electric furnace at 640°C. Next, 2%wt Ca was added to melt followed by addition of 1 % wt $TiH_2$ and mixing for a few minutes in the crucible. The mixture containing bubbles' nuclei was cast in a pre-heated steel mold and transferred to a furnace kept at 680°C. Finally, the mold was quenched in water and solidified. The solid foam was cut for metallography testes. Results were compared with virtual images from modeling tries.

## 4. Results and Discussion

Simulation results for two bubbles merging neglecting gravity are shown in Fig. 2. Three stages are indicated: approach, contact point formation and film formation. These plots show the variation of fluid pressure and velocity between bubbles. Despite common models, merging did not happen, Fig. 2b. Film rapture starts when $\partial^2 p/\partial n^2=0$. Results show that merging is function of bubbles' radius. Simulation results for 9 bubble sizes in A356 alloy are depicted in Fig. 3. Microstructures are compared in Fig. 4 for real and simulated parts.

Both the LB method by unmodified Shan-Chen model in OpenLB and FEM method with level set model in COMSOL software yield to the same result (merging bubbles at contact point without film formation). However, this behavior is not acceptable for molten metals foaming process since there is no kiss-point at the intersection of two bubbles. Therefore, none of the existing methods works well. The modified Shan-Chen method developed here is expected to simulate the situation more accurately.

Results, Fig. 2, indicate that a film is formed at the interface. No merging occurs until a specific criterion is met. This confirms the capability of the proposed model in simulating real conditions.

The pressure and velocity profiles across a horizontal line between bubbles are presented in Fig. 2. As the film thickens, an instability occurs in a specific thickness (which is loosely dependent on bubble size). This instability in the pressure profile is similar to the Spinodal decomposition (Fig. 2b). Thus, the second derivative of pressure in the lattice point would be zero, indicating the initialization of wall rupture. So it is possible to determine the wall rupture using the pressure profile and processing its second derivative instead of detecting the critical thickness. By applying this criterion to the domain of foam formation, merging conditions for bubbles would be obtainable at each lattice point, enabling us to model metal foam formation process with better accuracy. The critical thickness ($h_c$) of the film is a function of discs radius of thin films ($R_d$) as $h_c = 0.002R_d^2 - 0.003R_d + 0.035$, and that of the bubble radius ($R_b$) as $R_d = 2.088\ln(R_b) + 0.114$, Fig. 3. In this model, film thickness increases with bubble size due to the faster growth of larger bubbles. Bubbles start to interact when they reach each other and simulation continues until the first cell ruptures. Continued simulation is accompanied by merging and coalescence, leading to drainage in metal foam structure.

In Fig. 4, simulation results are compared with real A356 foam samples. The accuracy of the simulation (evaluated by PPI, bubble area and morphology, and number of merging defects) is above 95%. Therefore the present code can simulate and predict bubble structure of A356 alloy before merging, drainage, and aging of liquid foam.

## 5. Conclusions

A new model based on Shan-Chen model was proposed in this work for simulation of metal foams at micro and meso scale during foaming stage. In contrast to the existing models, the new model is capable of simulation multi-bubbles with no limitation of number of bubbles. An accuracy better than 95% was confirmed when comparing simulation results with experimental observations in A356 alloy. The model can consider effect of the attraction-repulsion barriers between bubbles due to the impurity. Therefore it can be used for controlling and prediction of foam optimum density.

## 7. Figures

Fig. 1: Schematic demonstration of simulation steps: 1) Initial state of computational lattice with two gas bubbles, 2) separating the second phase grid, 3,4,5) placing each bubble in its own lattice and fluid in a separate lattice, 6) computing the velocities across the grid which leads to the formation of the coronas, and then coupling the lattices and calculating the interactions, 7) computing the new bubbles' size and updating the fluid domain, 8,9) merging lattices to postprocess the final output

Fig. 2: Pressure and velocity profiles across a horizontal line between two bubbles. Left: before contacting, Middle: contact point, Right: thin film formation.

Fig. 3: Relation of bubbles ($R_b$), disc ($R_d$) radius, and thickness of the thin film ($h_c$).

Fig. 4: Comparison between the real and the simulated structure of porous A356 Al foam by adding present code

# Figure 1
[Click here to download high resolution image](#)

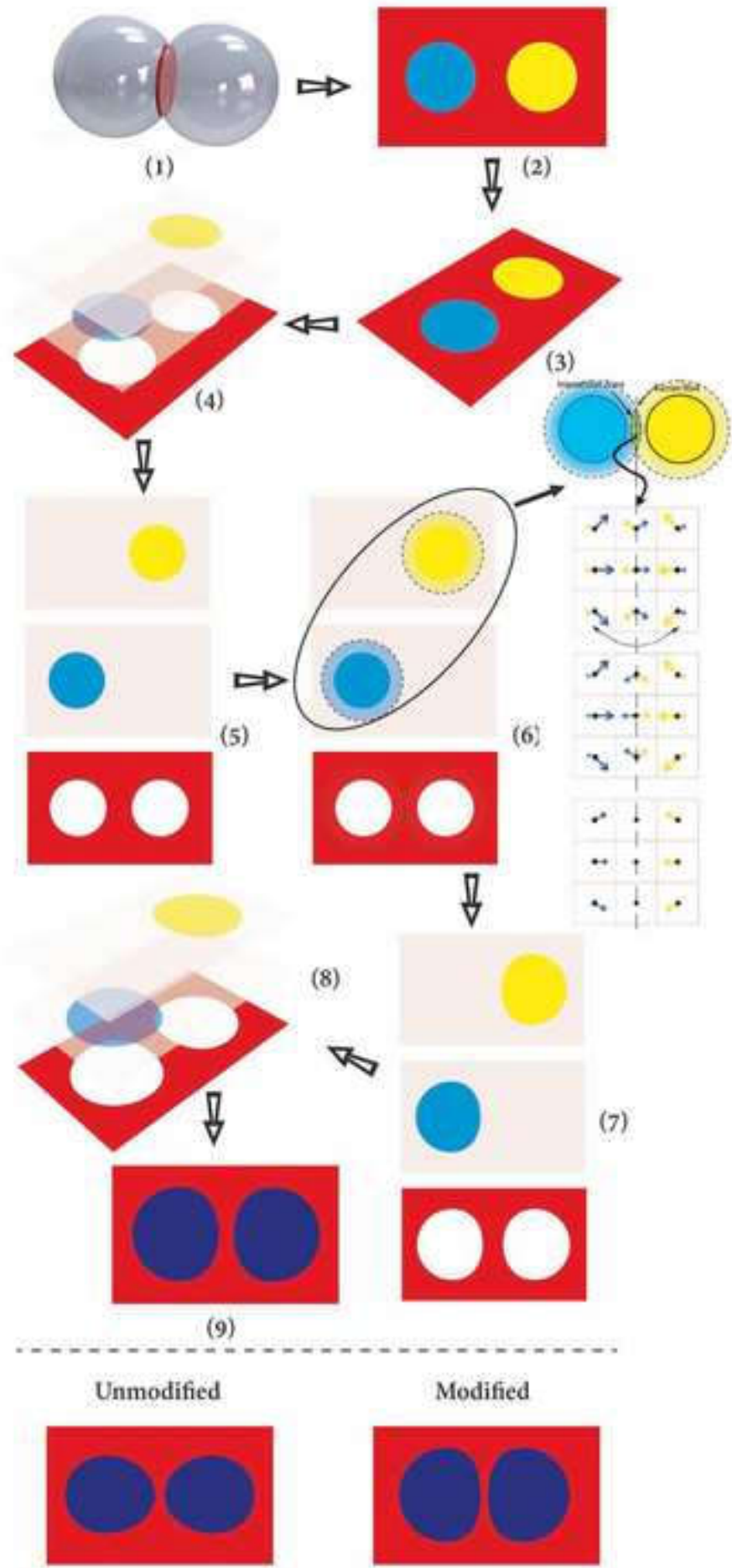

**Figure 2**
Click here to download high resolution image

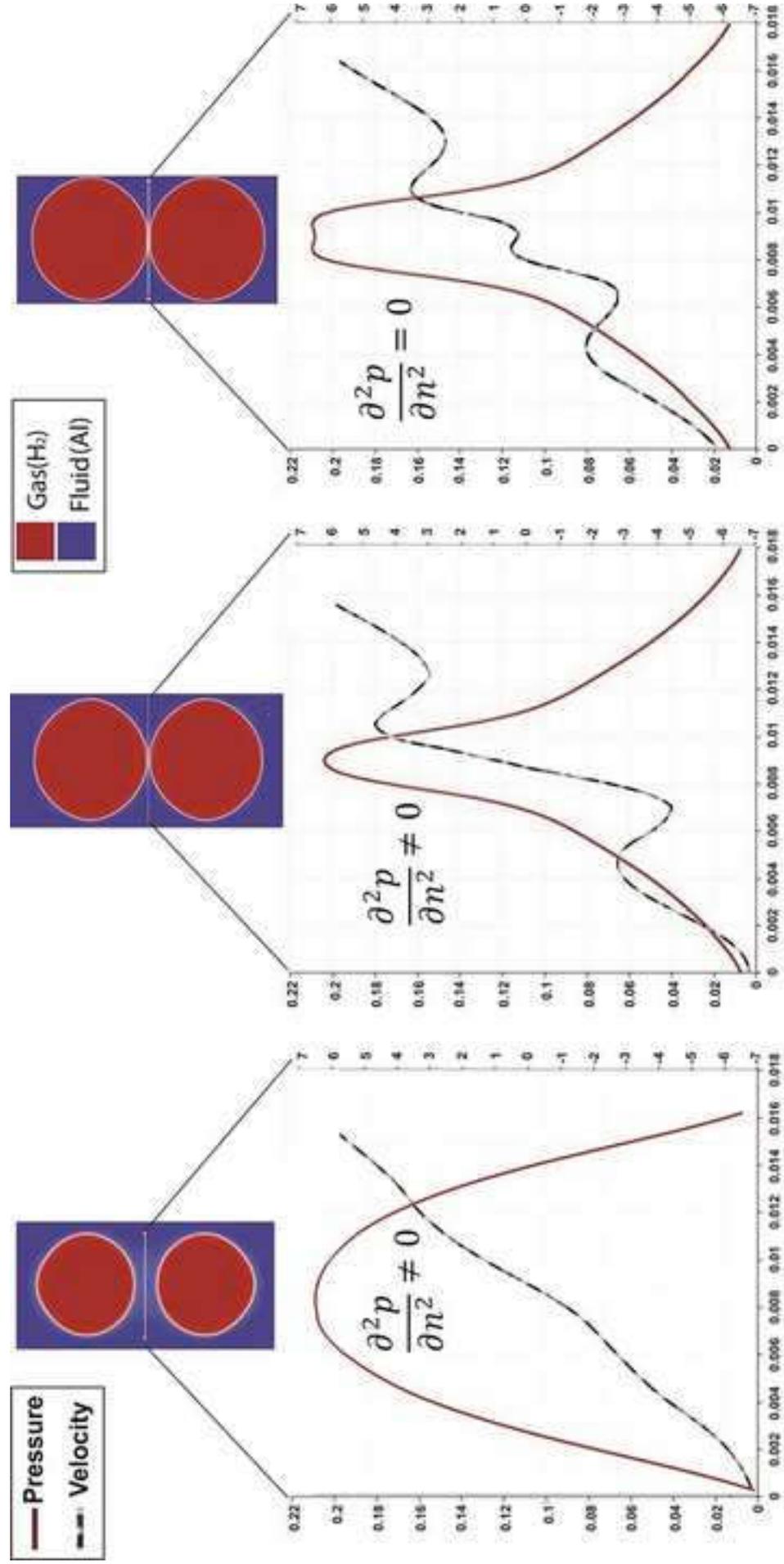

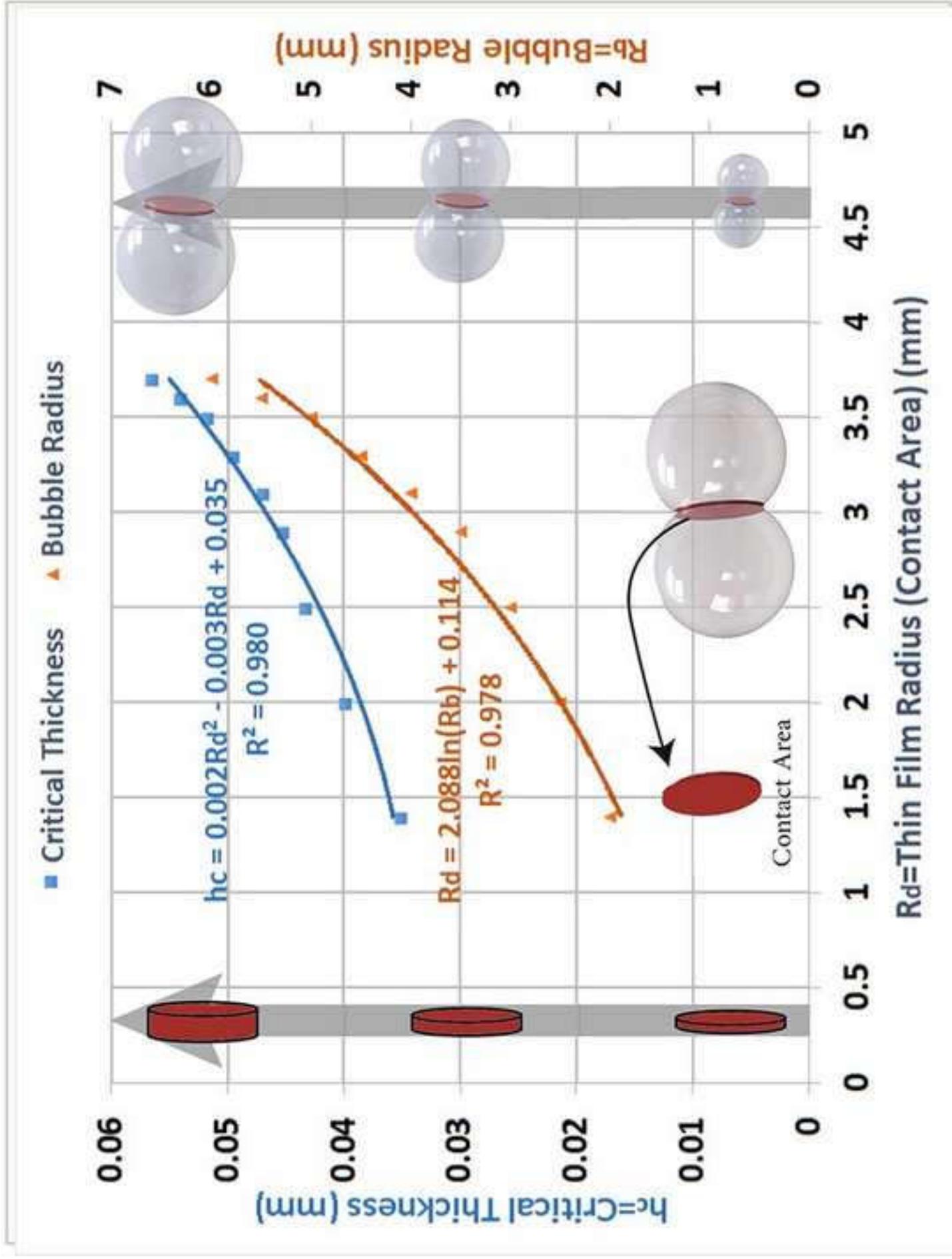

**Figure 3**
Click here to download high resolution image

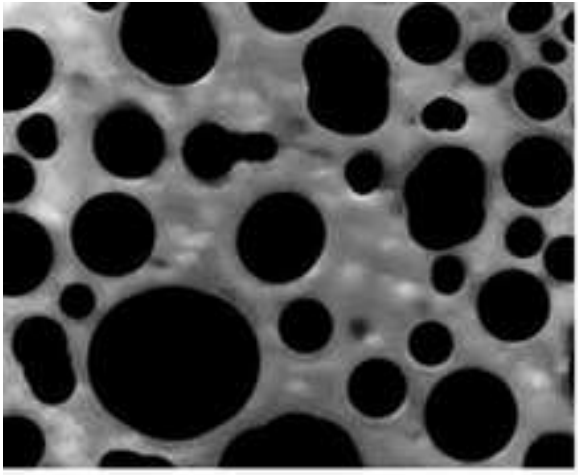 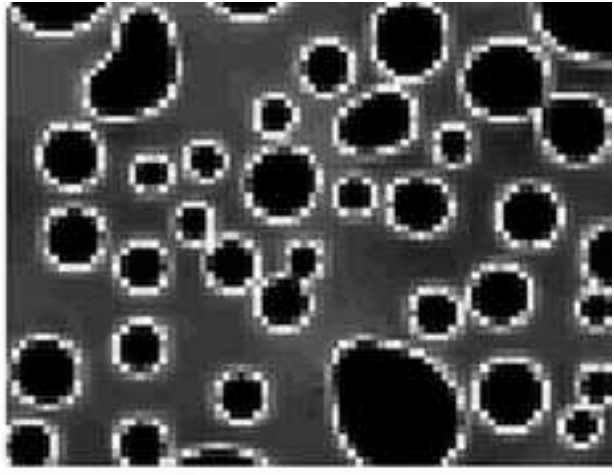

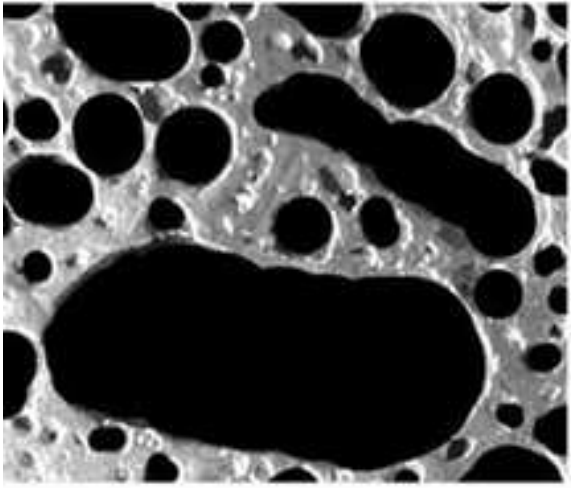 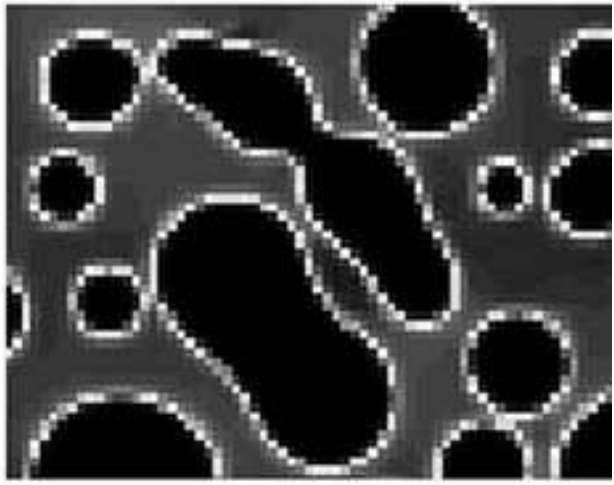

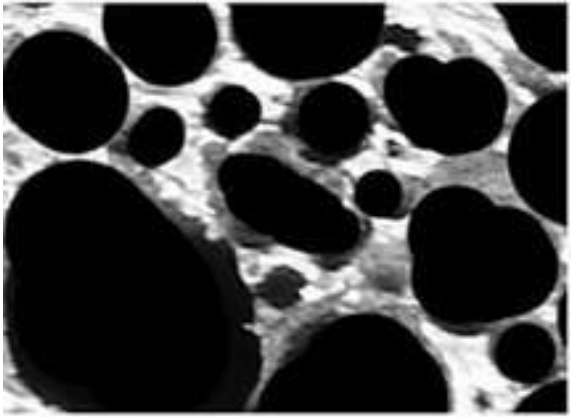 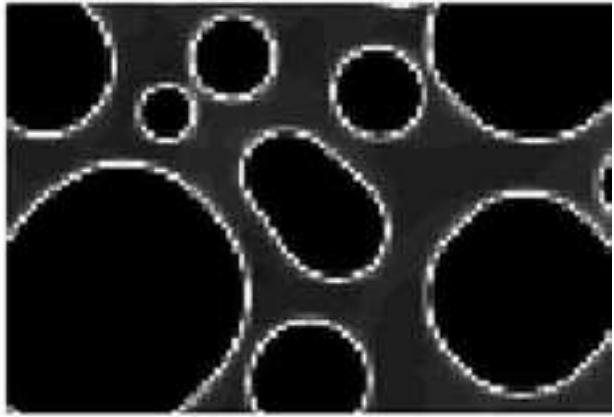

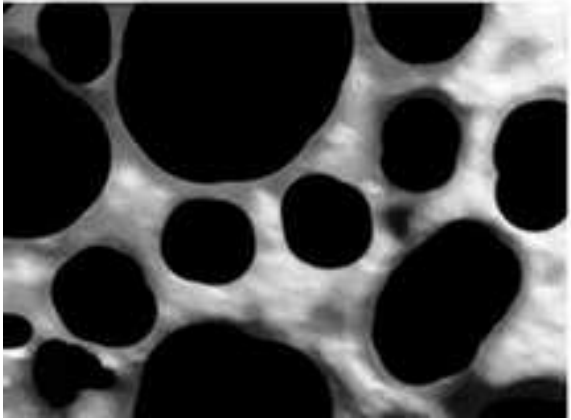 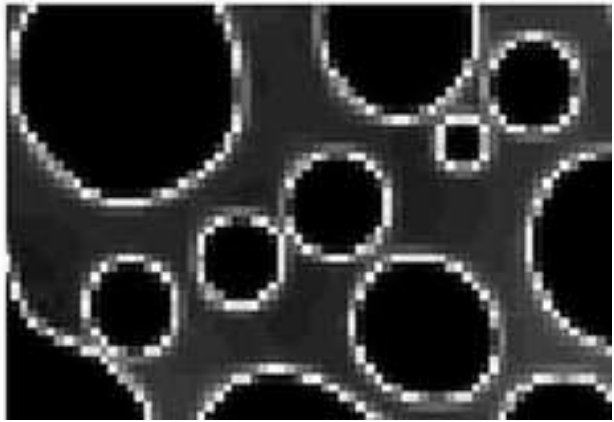